\documentclass[twocolumn]{autart}                                     

\usepackage{amsmath,amssymb}
\usepackage{enumerate}
\usepackage{chemarrow}
\usepackage{extarrows}
\usepackage{indentfirst}
\usepackage{cite}
\usepackage{graphicx}
\usepackage{setspace}
\usepackage{subeqnarray}
\usepackage{cases}             
\newtheorem{definition}{Definition}

\newtheorem{lemmax}{Lemma}

\newtheorem{remark}{Remark}
\newtheorem{example}{Example}

\begin{document}

\begin{frontmatter}

\title{Distributed Controller-Estimator for
Target Tracking of Networked Robotic Systems
under Sampled Interaction\thanksref{footnoteinfo}} 

\thanks[footnoteinfo]{This work was supported in part by
the National Natural Science Foundation of China under
Grant 61473128 and  the 973 National Basis Research Program
of China under Grant 2011CB013301.}

\author[Hust]{Ming-Feng Ge}\ead{fmgabc@163.com},    
\author[Hust]{Zhi-Hong Guan$^{*}$}\ead{zhguan@mail.hust.edu.cn},
\author[Hust]{Bin Hu},
\author[Hust]{Ding-Xin He},              
\author[Yangtze]{Rui-Quan Liao}  

\address[Hust]{College of Automation, Huazhong University of Science and Technology,
Wuhan, 430074, P. R. China}  
\address[Yangtze]{Petroleum Engineering College, Yangtze University, Jingzhou, 434023, P. R. China}        


\begin{abstract}                          
This paper investigates the target tracking problem for
networked robotic systems (NRSs) under sampled interaction.
The target is assumed to be time-varying and described by a second-order oscillator.
Two novel distributed controller-estimator algorithms (DCEA),
which consist of both continuous and discontinuous signals, are presented.
Based on the properties of small-value norms and Lyapunov stability theory,
the conditions on the interaction topology, the sampling period,
and the other control parameters are given such that
the practical stability of the tracking error is achieved
and the stability region is regulated quantitatively.
The advantages of the presented DCEA are illustrated
by comparisons with each other and the existing coordination algorithms.
Simulation examples are given to demonstrate the theoretical results.
\end{abstract}

\end{frontmatter}

\setlength{\parindent}{1em}
\setlength{\parskip}{1\baselineskip}

\section{Introduction}

Numerous real-world applications have required a group of interconnected robots
(networked robotic system, NRS) to accomplish one or several global tasks cooperatively, for instance,
assembly of heterogeneous robots, trajectory tracking of networked mobile robots,
operation of multi-fingered hands, management of intelligent highways {\cite{fmg01,fmg02,fmg03,fmg04,fmg05}}.
Distributed algorithms have been widely invoked in these applications due to their advantages,
including strong robustness, low consumptions and high efficiency {\cite{Olfati,Ren01}}.
Distributed consensus for NRSs under continuous interaction has been studied in {\cite{Liu01,Wang01}}.
Consensus tracking of a constant value of NRSs under undirected continuous interaction
has been investigated in {\cite{Dong01}}.
Consensus tracking of time-varying trajectories of NRSs under continuous interaction
has been studied in {\cite{Khoo01,Mei01}}.
Note that the aforementioned literatures have focused on continuous interaction,
which leads to higher interaction cost comparing with sampled interaction.
\setlength{\parskip}{0\baselineskip}
\par
Since sampling operation is one of the inevitable steps to realize digital interaction
in practical applications, the effect of sampled interaction for networked systems has been well studied recently,
see {\cite{Cao09,Gao09}} and references therein.
Impulsive control is an efficient technology in coordination of networked systems under sampled interaction {\cite{Guan02,Guan01}}.
Comparing with continuous control, impulsive control has shown satisfying performance
and prominent superiorities, including faster transient, less cost, lower computation, more flexible design {\cite{Guan03,Guan04,GHS}}.
However, the existing schemes cannot be directly applied to NRSs due to their inherent characteristics,
including strong nonlinearity, tight coupling, complex construction and fragility to chattering.
\par
Motivated by the above discussions,
two novel DCEA, consist of PD-like controllers and sampled-data estimators,
are given for target tracking of NRSs under sampled interaction.
The main contributions are summarized as follows.
1) Comparing with the coordination algorithms for NRSs under
continuous interaction {\cite{Wang01,Dong01,Khoo01,Mei01}},
we focus on sampled interaction, which can prominently reduce the interaction cost.
2) Comparing with the sampled-data coordination algorithms for single- and
double-integrator networks with a constant agreement value {\cite{Cao09,Gao09,Guan01,Guan02,Guan03,Guan04}},
we develop sampled-data coordination algorithms for NRSs with a time-varying target.
3) The presented DCEA provide a theoretical guidance for sampled-data control and estimation
of many physical networks with complex and strong nonlinear dynamics.

\textsl{Notations:}
$\mathbb{Z}^{\dag}$, $\mathbb{C}$ and $\mathbb R$ are the sets of positive integers,
complex numbers and real numbers, respectively.
$\left\| \cdot \right\|_1$, $\left\| \cdot \right\|_2$ and $\left\| \cdot \right\|_\infty$
are the $1$-norm, the $2$-norm and the $\infty$-norm, respectively.
${\mathbf 1} = {\rm{col}} [ 1, \cdots, 1 ]$ and ${\mathbf 0} = {\rm{col}} [ 0, \cdots, 0 ]$
are the column vectors of proper dimensions, respectively.
$I_p$ is the identity matrix of order $p$.
${\rm Re}(\cdot)$ and ${\rm Im}(\cdot)$ are the real part and the imaginary part of a complex number, respectively.
$\lambda_{\min} ( \cdot )$, $\lambda_{\max} ( \cdot )$, $\sigma(\cdot)$ and $\det(\cdot)$
denote the minimum eigenvalue, the maximum eigenvalue, the spectrum and the determinant of a matrix, respectively.
We say that $\mathcal{A} \in \Omega^{p \times p}$ if the eigenvalues of $\mathcal{A}$
lie within the open unit disc, $\forall \mathcal{A} \in {\mathbb C}^{p \times p}$.
We say that $\eta_1 \in \mathcal{U}\left( {\eta_2 ;\delta } \right)$ if and only if
$\left\| {\eta_1 - \eta_2 } \right\|_2 \leq \delta$, $\forall \delta > 0, \eta_1,\eta_2 \in {\mathbb R}^p$.

\setlength{\parskip}{1\baselineskip}

\section{Problem Formulation and Preliminaries}

\subsection{\sc Dynamic Models and Control Problem}

Following \cite{Lewis}, a NRS is described by
\begin{equation}\label{0.1}
{{\mathcal M}_i}\left( q_i \right){{\ddot q}_i} + {{\mathcal C}_i}\left( {{{\dot q}_i},{q_i}} \right){{\dot q}_i}
+ {{\mathcal G}_i}\left( q_i \right) = {\tau _i} + {\tau _{id}},
\end{equation}
where $q_i(t)$, $\tau _i(t)$ and ${\tau _{id}}(t)$ are abbreviated to $q_i$, $\tau _i$ and ${\tau _{id}}$, respectively,
$t \in \mathcal Q = \left[ {{t_0}, + \infty } \right)$, ${t_0} \geq 0$ is the initial time,
$i \in \mathcal{I} = \{ 1, \cdots ,n \}$,
$q_i,\dot q_i,\ddot q_i \in {\mathbb R}^m$ are the vectors of position, velocity and acceleration, respectively,
${{\mathcal{M}}_i}\left( q_i \right),{\mathcal{C}_i}({\dot q_i},{q_i}) \in {\mathbb R}^{m \times m}$
are the inertia matrix and the centripetal-Coriolis matrix, respectively,
${{\mathcal G}_i}\left( {q_i} \right),{\tau _i},{\tau _{id}} \in {\mathbb R}^m$
are the vectors of gravity, input and disturbance, respectively.
\par
\setlength{\parskip}{0\baselineskip}
The target is described by a second-order oscillator
\begin{equation*}
  {\dot \varepsilon_0}(t) = {\upsilon_0}(t), ~
  {\dot \upsilon_0}(t) = {a_0}(t),
\end{equation*}
where $\varepsilon_0,\upsilon_0,a_0 \in {\mathbb R}^m$
denote the vectors of position, velocity and acceleration, respectively.
\par
The control problem is to develop proper input ${\tau _i}$ for target tracking of
robot $i (\forall i \in {\mathcal I})$ with its own states and
the sampled data of its neighbours,
$i.e.$, for any $i \in {\mathcal I}$,
\begin{equation*}
\mathop {\lim }\limits_{t \to \infty } {q_i}(t)  \in {\mathcal U}\left( {{\varepsilon_0};{\rho_1}} \right),~
\mathop {\lim }\limits_{t \to \infty } {\dot{q}_i}(t)  \in {\mathcal U}\left( {{\upsilon_0};{\rho_2}} \right),
\end{equation*}
where $\rho _1,\rho_2 > 0$ can be sufficiently small
by choosing appropriate parameters for the DCEA designed later.
\par
The Euler-Lagrange system (\ref{0.1}) is a suitable
description for many NRSs, including multiple manipulators,
multi-fingered hands and networked mobile robots, to name a few {\cite{Kelly}}.
Then the following properties for Euler-Lagrange systems are introduced {\cite{Feng02}}.
\par
\begin{enumerate}[(\emph{P}1)]
  \item ${{\mathcal M}_i}\left( q_i \right)$ is symmetric and positive definite;\label{p1}
  \item ${{\dot {\mathcal M}}_i}({q_i})- 2{{\mathcal C}_i}({\dot q_i},{q_i})$ is skew-symmetric;\label{p2}
  \item $\lambda_{im} \leq \left\| {\mathcal M}_i(q_i) \right\|_2 \leq \lambda_{iM}$,
    $\left\| {\mathcal C}_i(\eta,{q_i}) \right\|_2 \leq \lambda_{ic} \left\| \eta \right\|_2 $,
    $\left\| {\mathcal G}_i(q_i) \right\|_2 \leq \lambda_{ig} $,
    $\left\| {\tau _{id}} \right\|_2 \leq \lambda_{id} $,
    $\forall \eta,{q_i} \in {{\mathbb R}^m}$,
    where $\lambda_{im},\lambda_{iM},\lambda_{ic},\lambda_{ig},\lambda_{id} > 0$ are positive constants.\label{p3}
\end{enumerate}
\setlength{\parskip}{1\baselineskip}
\subsection{\sc Graph Theory and Lemmas}

Let $\mathcal{J} = \{ 0,1,\cdots,n \} \supset {\mathcal I}$,
where node $0$ is the target, node $i$ is robot $i$.
The NRS interaction is denoted by a digraph ${\Im} = \{ \mathcal{J}, \mathcal{E}, \mathcal{W} \}$
with edge set $\mathcal{E} \subseteq \mathcal{J} \times \mathcal{J}$.
An edge $\{ j,i \} \in \mathcal{E}$ means node $i$ can access the information of node $j$ directly,
but not vice versa.
$\mathcal{W} = [ w_{ij} ]_{(n + 1) \times (n + 1)}$ is the adjacency matrix,
where $w_{ii} = 0$; $w_{ij} > 0 \Leftrightarrow \{ j,i \} \in \mathcal{E}$;
$w_{ij} = 0$ otherwise, $\forall i,j \in \mathcal J$.
A directed path is a finite ordered sequence $\{ i_1,i_2 \}, \{ i_2,i_3 \}, \cdots$, in a digraph.
A digraph contains a spanning tree means that there exists a root node that has a directed path to the other nodes.
Node $0$ (the target) is the root node.
Let $\hat{\mathcal W} = [ w_{ij} ]_{n \times n}$,
$\zeta = {\rm col} (w_{10},\cdots,w_{n0})$,
$\varpi_i = {\sum\nolimits_{\iota \in \mathcal{J}} {w_{i\iota}} }$,
$\mathcal{B} = {\rm{diag}}( \varpi_1, \cdots ,\varpi_n )$,
$\mathcal{D} = \mathcal{B}^{-1} \hat{\mathcal W}$, $\forall i,j \in {\mathcal I}$.
\par
\setlength{\parskip}{0\baselineskip}
The NRS interaction only occurs at the sampling time $t_k$
and is thus called sampled interaction.
The sampling time sequence $\left\{ t_1, \cdots, t_k, \cdots \right\}$ satisfies
$t_0 < t_{1}<\cdots <t_{k}<\cdots$, $\mathop {\lim }\nolimits_{k\to +\infty }t_{k}= + \infty$,
$t_k - t_{k-1} = h~(\forall k \in \mathbb{Z}^\dag)$,
where $h > 0$ is the sampling period.
Then the following assumptions are made.
\begin{enumerate}[(\emph{A}1)]
  \item $\left\| {{v_0}(t)} \right\|_\infty \leq {\gamma _1}$,
    $\left\| {{a_0}(t)} \right\|_\infty \leq \gamma _2$, $\forall t \in {\mathcal Q}$,
    where ${\gamma _1},{\gamma _2} > 0$ are positive constants;\label{a2}
  \item The digraph $\Im$ contains a spanning tree at
    each sampling time $t_k$, $\forall k \in \mathbb{Z}^\dag$.\label{a1}
\end{enumerate}
\setlength{\parskip}{1\baselineskip}
\begin{definition}
For any $\mathcal{A} \in \Omega^{p \times p}$, a matrix norm $\left\| \cdot \right\|_\mathcal{A}$
is called small-value norm if $\left\| \mathcal{A} \right\|_\mathcal{A} < 1$.
Additionally, a vector norm $\left\| \cdot \right\|_\mathcal{A}$ defined as
$\left\| \eta \right\|_\mathcal{A} = \left\| \eta {\bf 1}^T \right\|_\mathcal{A}(\forall \eta \in {\mathbb R}^p)$
is also called small-value norm.
\end{definition}

\begin{lemmax}\label{l1}
{\cite{Cao09}}
Suppose that Assumption A{\ref{a1}} holds.
$\mathcal{B}$ is invertible,
$ \left\| {\mathcal D}^n \right\|_\infty < 1$,
$i.e.$, ${\mathcal D} \in \Omega^{n \times n}$.
\end{lemmax}

\begin{lemmax}\label{l3}
For any $\mathcal{A} \in \Omega^{p \times p}$, there exist corresponding
small-value norms $\left\| \cdot \right\|_\mathcal{A}$,
including a matrix norm and a vector norm.
Let $\eta_1,\eta_2 \in {\mathbb R}^p$ and $\mathcal{H} \in {\mathbb R}^{p \times p}$.
Then $\left\| \cdot \right\|_\mathcal{A}$ satisfies:
1) $\left\| \mathcal{H}\eta_1 \right\|_\mathcal{A} \leq \left\| \mathcal{H} \right\|_\mathcal{A} \left\| \eta_1 \right\|_\mathcal{A}$;
2) $\left\| \eta_1 + \eta_2 \right\|_\mathcal{A} \leq \left\| \eta_1 \right\|_\mathcal{A} + \left\| \eta_2 \right\|_\mathcal{A}$;
3) the vector norm $\left\| \cdot \right\|_\mathcal{A}$ is equivalent to $\left\| \cdot \right\|_2$ and $\left\| \cdot \right\|_\infty$.
\end{lemmax}

\section{Target Tracking over Sampled Interaction}

\subsection{\sc Distributed Controller-Estimator Algorithms}

Let ${\varepsilon}_i$ and ${\upsilon}_i$ be the estimate value of
${\varepsilon}_0$ and ${\upsilon}_0$ for robot $i$, $\forall i \in \mathcal{I}$.
The first-order DCEA is given as
\begin{equation}\label{3.3}
\left\{ \begin{array}{lll}
   {\tau _i} = \mathcal{K}_{ip} ( {{\varepsilon} _i} - {q _i} )
         + \mathcal{K}_{id} ( {{\upsilon} _i} - {\dot{q} _i} ), \\
   {\dot {\varepsilon} _i}  =  {\upsilon }_i,~~~{\dot {\upsilon } _i} = 0, ~~~~~~
   t \in \left( {t_{k - 1}},{t_k} \right], \\
    \Delta  {\varepsilon} _i = \alpha \sum\limits_{j \in \mathcal J} \frac{w_{ij}}{\varpi_i} \left( {\varepsilon} _j - {\varepsilon} _i \right), ~~~ t = {t_k}, \\
    \Delta  {\upsilon} _i = \beta \sum\limits_{j \in \mathcal J} \frac{w_{ij}}{\varpi_i} \left( {\upsilon} _j - {\upsilon} _i \right), ~~ t = {t_k},
   \end{array} \right.
\end{equation}
and the second-order DCEA is given as
\begin{equation}\label{2.2}
\left\{ \begin{array}{lll}
   {\tau _i} = \mathcal{K}_{ip}( {{\varepsilon} _i} - {q _i} )
         + \mathcal{K}_{id} ( {{\upsilon} _i} - {\dot{q} _i} \big), \\
   {\dot {\varepsilon} _i} = {\upsilon }_i,~~~{\dot {\upsilon } _i} = 0, ~~~~~~~~
   t \in \left( {t_{k - 1}},{t_k} \right], \\
    \Delta {\upsilon} _i = \alpha \sum\limits_{j \in \mathcal J} \frac{w_{ij}}{\varpi_i} \left( {\varepsilon} _j - {\varepsilon} _i \right) \\
    ~~~~~~~~~ + \beta \sum\limits_{j \in \mathcal J} \frac{w_{ij}}{\varpi_i} \left( {\upsilon} _j - {\upsilon} _i \right), ~~ t = {t_k},
   \end{array} \right.
\end{equation}
where $\alpha, \beta > 0$,
$\mathcal{K}_{ip},\mathcal{K}_{id}$ are positive definite matrices,
$\Delta {\varepsilon} _i ( t_k ) = { {\varepsilon} _i} ( t_k^ + ) - {{\varepsilon} _i} ( t_k )$,
${ {\varepsilon} _i} ( t_k^ + ) = \mathop {\lim }_{\sigma \to 0^+} {\varepsilon} _i ( t_k + \sigma )$,
and $\Delta {\upsilon } _i ( t_k ) $ is defined analogously.
It is assumed that ${\varepsilon} _i \left( t \right)$ and ${\upsilon} _i \left( t \right)$ are left continuous at $t = {t_k}$, $\forall k \in \mathbb{Z}^\dag$.

\begin{remark}
Note that ${\varepsilon} _i$ and ${\upsilon} _i$ jump at each sampling time instant $t_k$
in the first-order DCEA, while only ${\upsilon} _i$ jumps in the second-order DCEA.
Thus, the trajectories of $q_i$ and ${\varepsilon} _i$ for the NRS under the
second-order DCEA are smoother than that under the first-order DCEA.
\end{remark}

\subsection{\sc Analysis of First-order Algorithm}

Substituting DCEA (\ref{3.3}) into NRS (\ref{0.1}) gives that
\begin{equation}\label{2.1}
\left\{ {\begin{array}{*{20}{lll}}
{{\mathcal M}_i}\left( {q_i} \right){{\ddot q}_i} + {{\mathcal C}_i}\left( {{{\dot q}_i},{q_i}} \right){{\dot q}_i} + {{\mathcal G}_i}\left( {{q_i}} \right) \\
~~~~~~~~~~ = \mathcal{K}_{ip} ( {{\varepsilon} _i} - {q _i} ) + \mathcal{K}_{id} ( {{\upsilon} _i} - {\dot{q} _i} ) + {\tau _{id}}, \\
{\varepsilon} _i \left( t_{k+1} \right) = {\varepsilon }_i(t_k^+) + h {\upsilon }_i(t_k^+), \\
{\upsilon } _i \left( t_{k+1} \right) = {\upsilon }_i(t_k^+), \\
{\varepsilon }_i(t_k^+) = {\varepsilon }_i(t_k) + \alpha \sum\limits_{j \in \mathcal J} \frac{w_{ij}}{\varpi_i} \left( {{\varepsilon} _j\left( t_k \right) - {\varepsilon} _i\left( t_k \right)} \right), \\
{\upsilon }_i(t_k^+) = {\upsilon }_i(t_k) +\beta \sum\limits_{j \in \mathcal J} \frac{w_{ij}}{\varpi_i} \left( {{\upsilon} _j\left( t_k \right) - {\upsilon} _i\left( t_k \right)} \right),
\end{array}} \right.
\end{equation}
where ${\varepsilon} _i \left( t_1 \right) = {\varepsilon} _i \left( t_0 \right) + h {\upsilon} _i \left( t_0 \right)$, ${\upsilon} _i \left( t_1 \right) = {\upsilon} _i \left( t_0 \right)$, $i \in {\mathcal I}$, $k \in \mathbb{Z}^\dag$.
\par
\setlength{\parskip}{0\baselineskip}
Let $e_i = {q _i} - {\varepsilon _0}$,
$\dot e_i = {\dot q _i} - {\upsilon _0}$,
$\bar{\varepsilon} _i = {{\varepsilon} _i} - {{\varepsilon}_0}$,
$\bar{\upsilon} _i = {{\upsilon} _i} - {{\upsilon}_0}$,
$\bar{\varepsilon} = {\rm col} \left( {\bar \varepsilon}_1, \cdots ,{\bar \varepsilon}_n \right)$,
$\bar{\upsilon} = {\rm col} \left( {\bar \upsilon}_1, \cdots ,{\bar \upsilon}_n \right)$,
$x = {\rm col} \left( {\bar \varepsilon},{\bar \upsilon} \right) $,
$\forall i \in {\mathcal I}$.
Then system (\ref{2.1}) becomes
\begin{equation}\label{2.3}
\left\{ {\begin{array}{*{20}{lll}}
{{\mathcal M}_i}\left( q_i \right){{\ddot e}_i}
+ {{\mathcal C}_i}\left( {{\dot q}_i},{q_i} \right) {{\dot e}_i}  = {\phi _i}(t) - \mathcal{K}_{ip} e_i - \mathcal{K}_{id} \dot e_i, \\
x (t_{k+1}) = \Lambda x (t_k) + \Delta(k),
\end{array}} \right.
\end{equation}
where $\Delta (k) = {\rm col} [ {\mathbf 1} \otimes \Delta_1(k), {\mathbf 1} \otimes \Delta_2(k) ]$,
$\otimes$ denotes the Kronecker product,
$\Delta_1 (k) = {{\varepsilon}_0} (t_k) - {{\varepsilon}_0} (t_{k+1}) + h { \upsilon }_0(t_k)$,
$\Delta_2 (k) = {{\upsilon}_0} (t_k) - {{\upsilon}_0} (t_{k+1})$,
${\phi _i}(t) = {\tau _{id}} - {{\mathcal M}_i}\left( q_i \right) {a_0} - {{\mathcal C}_i}\left( {{\dot q}_i},{q_i} \right) {v_0} - {{\mathcal G}_i}\left( q_i \right) + \mathcal{K}_{ip} \bar{\varepsilon} _i\left( t \right) + \mathcal{K}_{id} \bar{\upsilon} _i\left( t \right)$,
\begin{align*}
    \Lambda = \left[ {\begin{array}{*{20}{ccc}}
    {(1-\alpha) I_{n} + \alpha {\mathcal D} }&{h [(1-\beta) I_{n} + \beta {\mathcal D} ]}\\
    {0}&{(1-\beta) I_{n} + \beta {\mathcal D} }
    \end{array}} \right] \otimes I_m.
\end{align*}
\par
Then the solution of the control problem is to analyze the asymptotic behaviors
of the states of system (\ref{2.3}).
\par
Let $s_i \in \mathbb{C}$ be the \emph{i-th} ($i =1,\cdots,n$) eigenvalue of $\mathcal D$,
\begin{equation}\label{0.5}
  \begin{array}{lll}
    \kappa_1 = b_1 c_1 \max(2 \gamma_1 ; \gamma_2) (1 - \left\| \Lambda \right\|_\Lambda)^{-1}, \\
    \kappa_2 = b_2 c_2 \gamma_2 (1 - \left\| {\mathcal P} \right\|_{\mathcal P})^{-1},
  \end{array}
\end{equation}
where ${\mathcal P} = (1-\beta) I_{mn} + \beta {\mathcal D} \otimes I_{m}$,
$\left\| \cdot \right\|_\Lambda$ and $\left\| \cdot \right\|_{\mathcal P}$
are small-value norms, $b_1,c_1,b_2,c_2$ are positive constants such that
$\left\| \eta \right\|_\infty \leq b_1 \left\| \eta \right\|_\Lambda$,
$\left\| \eta \right\|_\Lambda \leq c_1 \left\| \eta \right\|_\infty$,
$\left\| \eta \right\|_\infty \leq b_2 \left\| \eta \right\|_{\mathcal P}$,
$\left\| \eta \right\|_{\mathcal P} \leq c_2 \left\| \eta \right\|_\infty$.
Then the convergence analysis of $\bar{\varepsilon} _i$ and $\bar{\upsilon} _i$ is given
in Theorem \ref{t1} proved in Appendix \ref{appendix1}.
\setlength{\parskip}{1\baselineskip}
\begin{thm}\label{t1}
Suppose that Assumptions A\ref{a2}-A\ref{a1} hold. If
\begin{equation}\label{0.8}
  0 < \alpha,\beta < \mathop {\min }\limits_{s_i \in \sigma({\mathcal D})} \frac{{2 - 2Re({s_i})}}{{{{\left| {1 - {s_i}} \right|}^2}}},
\end{equation}
then DCEA (\ref{3.3}) implies that for any $i \in {\mathcal I}$,
\begin{equation}\label{6.6}
\left\{ \begin{array}{lll}
  \mathop {\lim }\limits_{t \to \infty } \left\| {\bar \varepsilon}_i \right\|_\infty \leq \delta_1, \\
  \mathop {\lim }\limits_{t \to \infty } \left\| {\bar \upsilon}_i \right\|_\infty \leq \delta_2, \\
  \delta_1 = h \kappa_1, \\
  \delta_2 = h \min (\kappa_1; \kappa_2),
  \end{array} \right.
\end{equation}
where $\left| \cdot \right|$ denotes the modulus,
$h$ is the sampling period, $\kappa_1$ and $\kappa_2$ are given by (\ref{0.5}).
\hfill $\blacksquare$
\end{thm}

\begin{remark}
$\delta_1$ and $\delta_2$ are proportional to the sampling period $h$,
which means $\delta_1$ and $\delta_2$ can be sufficiently small by choosing $h$ small enough
and the formulas of $\left\| \cdot \right\|_\Lambda$, $\left\| \cdot \right\|_{\mathcal P}$
are not necessarily known for regulating $\delta_1$ and $\delta_2$.
\end{remark}
\par
Next based on Theorem \ref{t1}, the convergence of system (\ref{2.3}) is studied.
Assumption A\ref{a2} and Property P\ref{p3} implies
\begin{equation}\label{1.9}
\left\| {\phi _i}(t) \right\|_2 \leq \mu_{1i} + \mu_{2i} \left\| \dot e_i \right\|_2 + {\mu _{3i}}(t),
~\forall i \in \mathcal I,
\end{equation}
where $\mu _{1i},\mu_{2i} > 0$ are positive constants,
${\mu _{3i}}\left( t \right) = {\lambda _{\max }}\left( {{\mathcal{K}_{ip}}} \right) {\left\| {\bar \varepsilon }_i \right\|_2 } + {\lambda _{\max }}\left( {{\mathcal{K}_{id}}} \right) \left\| {\bar \upsilon }_i \right\|_2$.
For the first subsystem of (\ref{2.3}), consider the Lyapunov function candidate
\begin{equation}\label{4.0}
    \begin{array}{lll}
        \mathcal{V}_i = \frac{1}{2} {\dot e}_i^T {{\mathcal M}_i}\left( q_i \right) {\dot e}_i + {\dot e}_i^T {{\mathcal M}_i}\left( q_i \right) \tanh \left( e_i \right) \\
        ~~~~~~~ + \frac{1}{2} e_i^T \mathcal{K}_{ip} e_i
        + \left\| \mathcal{K}_{id} \ln (\cosh ({e_i})) \right\|_1,
    \end{array}
\end{equation}
where $\tanh ( \cdot ), \ln ( \cdot ), \cosh ( \cdot )$ are the function vectors of
hyperbolic tangent, natural logarithm and hyperbolic cosine, respectively.
Note that
\begin{equation*}
   \begin{array}{lll}
     \frac{1}{4} {\dot e}_i^T {{\mathcal M}_i}\left( q_i \right) {\dot e}_i
     + {\dot e}_i^T {{\mathcal M}_i}\left( q_i \right) \tanh \left( e_i \right)
     + \frac{1}{2} e_i^T \mathcal{K}_{ip} e_i  \\
     = \frac{1}{4} \big( {\dot e}_i + 2 \tanh \left( e_i \right) \big)^T {{\mathcal M}_i}\left( q_i \right) \big( {\dot e}_i + 2 \tanh \left( e_i \right) \big) \\
     ~~~ + \frac{1}{2} e_i^T \mathcal{K}_{ip} e_i - \tanh^T \left( e_i \right)  {{\mathcal M}_i}\left( q_i \right) \tanh \left( e_i \right) \\
     \geq \frac{1}{2} \big( \lambda_{\min} \left( \mathcal{K}_{ip} \right) - 2 \lambda_{iM} \big) e_i^T  e_i,
   \end{array}
\end{equation*}
where $\lambda_{iM}$ is given in Property P\ref{p3}.
Then by (\ref{4.0}),
\begin{equation*}
    \begin{array}{lll}
        \mathcal{V}_i \left( e_i,{\dot e}_i \right)
        &\geq& \frac{1}{4} {\dot e}_i^T {{\mathcal M}_i}\left( q_i \right) {\dot e}_i +
        \left\| \mathcal{K}_{id}~\ln (\cosh ({e_i})) \right\|_1 \\
        && + \frac{1}{2} \big( \lambda_{\min} \left( \mathcal{K}_{ip} \right) - 2 \lambda_{iM} \big) e_i^T  e_i.
    \end{array}
\end{equation*}
Thus $\lambda_{\min} \left( \mathcal{K}_{ip} \right) \geq 2 \lambda_{iM}$ means that
$\mathcal{V}_i$ is positive definite.
By (\ref{2.3}) and Properties P\ref{p1}-P\ref{p2}, the derivative of $\mathcal{V}_i$ is
\begin{equation*}
   \begin{array}{lll}
     {\dot {\mathcal V}_i}
     = {\phi _i^T}(t) \left( \dot e_i + \tanh \left( e_i \right) \right) + \dot e_i^T{\mathcal{M}_i}\left( {{q_i}} \right){{\rm{sech}}^2}\left( {{e_i}} \right){{\dot e}_i} \\
     ~ + \dot e_i^T{\mathcal{C}_i}\left( {{{\dot q}_i},{q_i}} \right)\tanh \left( {{e_i}} \right) - e_i^T{{\mathcal K}_{ip}}\tanh \left( {{e_i}} \right) - \dot e_i^T{{\mathcal K}_{id}}{{\dot e}_i},
   \end{array}
\end{equation*}
where ${\rm sech} ( \cdot )$ is the function vector of hyperbolic secant.
Since $\left\| \tanh \left( {e_i} \right) \right\|_\infty \leq 1$ and
$\lambda_{\max} \left( {\rm sech} \left( e_i \right) \right) = 1$, by (\ref{1.9}),
\begin{equation}\label{1.8}
    \begin{array}{lll}
        {\dot {\mathcal V}_i} \leq
        - [ \left( {{\lambda _{\min }}\left( {{{\mathcal K}_{id}}} \right) - {\varrho _{2i}}} \right){{\left\| {{{\dot e}_i}} \right\|}_2} - {\varrho _{3i}}(t) ] {\left\| {{{\dot e}_i}} \right\|_2}
        \\
        ~~~~~~~ - [ \left( {{\lambda _{\min }}\left( {{{\mathcal K}_{ip}}} \right) - {\varrho _{1i}}} \right){{\left\| {{e_i}} \right\|}_2} - {\varrho _{3i}}(t)] \\
        ~~~~~~~ \times {\left\| {\tanh \left( {{e_i}} \right)} \right\|_2},
    \end{array}
\end{equation}
where ${\varrho _{1i}},{\varrho _{2i}} > 0$ can be easily computed from (\ref{1.9}),
${\varrho _{3i}}(t) = {\mu _{1i}} + {\mu _{3i}}( t )$.
\par
\begin{thm}\label{t2}
Suppose that Assumptions A\ref{a2}-A\ref{a1} hold.
Using DCEA (\ref{3.3}) for (\ref{0.1}),
if ${\lambda _{\min }}\left( {{\mathcal{K}_{ip}}} \right) > \max \left( 2 \lambda_{iM} ; \varrho _{1i} \right)$, ${\lambda _{\min }}\left( {{{\mathcal K}_{id}}} \right) > \varrho _{2i}$ and (\ref{0.8}) hold,
then the control problem in this paper is solved, $i.e.$,
for any $i \in {\mathcal I}$,
\begin{equation}\label{1.7}
   \left\{ \begin{array}{lll}
        \mathop {\lim }\limits_{t \to \infty } {q_i} \in {\mathcal U}\left( {{\varepsilon_0};{\delta _3}} \right),~
        \mathop {\lim }\limits_{t \to \infty } {\dot{q}_i} \in {\mathcal U}\left( {{\upsilon_0};{\delta _4}} \right), \\
        \delta_3 = \mathop {\max} \limits_{i \in \mathcal I} \frac{\mu_{1i} + \sqrt m \left[ \delta _1 {\lambda _{\max }}\left( {{{\mathcal K}_{ip}}} \right) + \delta _2 {\lambda _{\max }}\left( {{{\mathcal K}_{id}}} \right) \right] } {{\lambda _{\min }}\left( {{{\mathcal K}_{ip}}} \right) - {\varrho _{1i}}}, \\
        \delta_4 = \mathop {\max} \limits_{i \in \mathcal I} \frac{ {\mu_{1i} + \sqrt m \left[ \delta _1 {\lambda _{\max }}\left( {{{\mathcal K}_{ip}}} \right) + \delta _2 {\lambda _{\max }}\left( {{{\mathcal K}_{id}}} \right) \right]} }{{\lambda _{\min }}\left( {{{\mathcal K}_{id}}} \right) - {\varrho _{2i}}},
    \end{array} \right.
\end{equation}
where $\delta _1$ and $\delta _2$ are presented in (\ref{6.6}).
\hfill $\blacksquare$
\end{thm}

{Proof.}\,\,
First $\lambda_{\min} \left( \mathcal{K}_{ip} \right) \geq 2 \lambda_{iM}$
ensures the positive-definiteness of $\mathcal{V}_i$.
Thus, if ${\lambda _{\min }}\left( {{{\mathcal K}_{ip}}} \right) > {\varrho _{1i}}$
and ${\lambda _{\min }}\left( {{{\mathcal K}_{id}}} \right) > {\varrho _{2i}}$, (\ref{1.8}) implies
\begin{equation*}
    \begin{array}{lll}
        \mathop {\lim }\limits_{t \to \infty } {\left\| {{e_i}\left( t \right)} \right\|_2} &\leq&
        \frac{\varrho _{3i}(t)}{{\lambda _{\min }}\left( {{{\mathcal K}_{ip}}} \right) - {\varrho _{1i}}}, \\
        \mathop {\lim }\limits_{t \to \infty } {\left\| {{{\dot e}_i}\left( t \right)} \right\|_2} &\leq&
        \frac{\varrho _{3i}(t)}{{\lambda _{\min }}\left( {{{\mathcal K}_{id}}} \right) - {\varrho _{2i}}}.
    \end{array}
\end{equation*}
By Theorem \ref{t1}, for any $i \in \mathcal I$,
\begin{equation*}
    \begin{array}{lll}
        ~~~ \mathop {\lim }\limits_{t \to \infty } {\mu _{3i}}\left( t \right) \\
        = {\lambda _{\max }}\left( {{{\mathcal K}_{ip}}} \right) \mathop {\lim }\limits_{t \to \infty } {\left\| {\bar \varepsilon }_i \right\|_2 }
        + {\lambda _{\max }}\left( {{{\mathcal K}_{id}}} \right) \mathop {\lim }\limits_{t \to \infty } {\left\| {\bar \upsilon }_i \right\|_2 } \\
        \leq h {\sqrt m} \left[ \kappa_1 {\lambda _{\max }}\left( {{{\mathcal K}_{ip}}} \right)
        + \min (\kappa_1;\kappa_2 ) {\lambda _{\max }}\left( {{{\mathcal K}_{id}}} \right) \right].
    \end{array}
\end{equation*}
Considering ${\varrho _{3i}}(t) = {\mu _{1i}} + {\mu _{3i}}\left( t \right)$,
we conclude that $\mathop {\lim }\nolimits_{t \to \infty } {\left\| e_i \right\|_2} \leq \delta_3$ and
$\mathop {\lim }\nolimits_{t \to \infty } {\left\| {\dot e}_i \right\|_2} \leq \delta_4$,
$\forall i \in {\mathcal I}$.
This completes the proof.
\hfill $\blacksquare$
\par
Suppose that (\ref{0.8}) holds.
By Theorem \ref{t2}, for any $\delta _3,\delta _4 > 0$,
if there exists a positive constant $\epsilon \in (0,1)$ such that
\begin{equation}\label{6.1}
 \left\{ \begin{array}{lll}
  \lambda _{\min }\left( {{{\mathcal K}_{ip}}} \right) \geq \max \left( 2 \lambda_{iM} ;  \frac{\mu_{1i}}{\epsilon\delta _3} + {\varrho _{1i}} \right), \\
  \lambda _{\min }\left( {{{\mathcal K}_{id}}} \right) \geq \frac{\mu_{1i}}{\epsilon\delta _4} + {\varrho _{2i}}, \\
  h \leq \frac{\mu_{1i} (1 - \epsilon) }
  {\epsilon\sqrt m \left[ \kappa _1 {\lambda _{\max }}\left( {{{\mathcal K}_{ip}}} \right) + \min (\kappa_1;\kappa_2 ) {\lambda _{\max }}\left( {{{\mathcal K}_{id}}} \right) \right]},
  \end{array} \right.
\end{equation}
then $\mathop {\lim }\nolimits_{t \to \infty } {q_i} \in {\mathcal U}\left( {{\varepsilon_0};{\delta _3}} \right)$ and
$\mathop {\lim }\nolimits_{t \to \infty } {\dot{q}_i} \in {\mathcal U}\left( {{\upsilon_0};{\delta _4}} \right).$

\begin{remark}\label{remark1}
(\ref{6.1}) means ${\mu _{1i}},{\mu _{2i}},{\varrho _{1i}},{\varrho _{2i}}$
are not necessarily known for regulating $\delta _3$ and $\delta _4$.
$\delta _3$ and $\delta _4$ can be sufficiently small by choosing
$\lambda _{\min }( {{{\mathcal K}_{ip}}} ),\lambda _{\min }( {{{\mathcal K}_{id}}} )$
large enough and $h$ small enough.
However, smaller $\lambda _{\min }\left( {{{\mathcal K}_{ip}}} \right)$ and
$\lambda _{\min }\left( {{{\mathcal K}_{id}}} \right)$ means $h$ can be larger,
the input cost and interaction consumption can be lower.
The parameter tuning strategy can be obtained by the trial-and-error methods \cite{SSF}.
\end{remark}

\subsection{\sc Analysis of Second-order Algorithm}

Substituting DCEA (\ref{2.2}) into (\ref{0.1}) gives
\begin{equation}\label{2.0}
\left\{ {\begin{array}{*{20}{lll}}
{{\mathcal M}_i}\left( q_i \right){{\ddot e}_i}
+ {{\mathcal C}_i}\left( {{\dot q}_i},{q_i} \right) {{\dot e}_i} + \mathcal{K}_{ip} e_i + \mathcal{K}_{id} \dot e_i = {\phi _i}, \\
x (t_{k+1}) = \Gamma x (t_k) + \Delta(k),
\end{array}} \right.
\end{equation}
where $x (t_1)$, $\Delta(k)$, ${\phi _i}(t)$ are given after (\ref{2.3}), and
\begin{equation*}
\Gamma = \left[ {\begin{array}{*{20}{ccc}}
    {(1 - \alpha h) I_{n} + \alpha h {\mathcal D} }&{h [(1-\beta) I_{n} + \beta {\mathcal D} ]}\\
    {-\alpha I_{n} + \alpha {\mathcal D}}&{(1-\beta) I_{n} + \beta {\mathcal D} }
    \end{array}} \right] \otimes I_m.
\end{equation*}
Let ${\theta _i} = {\mathop{\rm Re}\nolimits} \left( 2/{[1 - s_i]} \right)$,
${\vartheta _i} = {\mathop{\rm Im}\nolimits} \left( 2/{[1 - s_i]} \right)$ and
$ \kappa_3 = b_3 c_3 \max(2 \gamma_1 , \gamma_2) (1 - \left\| \Gamma \right\|_\Gamma)^{-1}$,
where $\left\| \cdot \right\|_\Gamma$ is the small-value norm,
$b_3,c_3 > 0$ satisfy
$\left\| \eta \right\|_\infty \leq b_3 \left\| \eta \right\|_\Gamma$,
$\left\| \eta \right\|_\Gamma \leq c_3 \left\| \eta \right\|_\infty$, $\forall \eta \in {\mathbb R}^{2mn}$.
\begin{thm}\label{t4}
Suppose that Assumptions A\ref{a2}-A\ref{a1} hold.
Using DCEA (\ref{2.2}) for (\ref{0.1}),
if ${\lambda _{\min }}\left( {{\mathcal{K}_{ip}}} \right) > \max \left( 2 \lambda_{iM} ; \varrho _{1i} \right)$,
${\lambda _{\min }}\left( {{{\mathcal K}_{id}}} \right) > \varrho _{2i}$, $\beta < \mathop {\min }\nolimits_{s_i \in \sigma({\mathcal D})} {\theta _i}$ and
\begin{equation}\label{5.2}
    0 < h < \mathop {\min }\limits_{s_i \in \sigma({\mathcal D})} \frac{{2{\beta ^2}\left( {{\theta _i} - \beta } \right)}}{{\alpha \left( {\vartheta _i^2 + {\beta ^2}} \right)}},
\end{equation}
then the control problem is solved, $i.e.$, for any $i \in {\mathcal I}$,
\begin{equation}\label{5.3}
    \left\{ \begin{array}{lll}
    \mathop {\lim }\limits_{t \to \infty } \left\| {\bar \varepsilon}_i \right\|_\infty \leq h \kappa_3,~\mathop {\lim }\limits_{t \to \infty } \left\| {\bar \upsilon}_i \right\|_\infty \leq h \kappa_3, \\
    \mathop {\lim }\limits_{t \to \infty } {q_i} \in {\mathcal U}\left( {{\varepsilon_0};{\delta _5}} \right),~
    \mathop {\lim }\limits_{t \to \infty } {\dot{q}_i} \in {\mathcal U}\left( {{\upsilon_0};{\delta _6}} \right), \\
    \delta _5 = \mathop {\max} \limits_{i \in \mathcal I} \frac{\mu_{1i} + h \kappa_3 \sqrt m \left[ {\lambda _{\max }}\left( {{{\mathcal K}_{ip}}} \right) + {\lambda _{\max }}\left( {{{\mathcal K}_{id}}} \right) \right]}{{\lambda _{\min }}\left( {{{\mathcal K}_{ip}}} \right) - {\varrho _{1i}}}, \\
    \delta _6 = \mathop {\max} \limits_{i \in \mathcal I} \frac{\mu_{1i} + h \kappa_3 \sqrt m \left[ {\lambda _{\max }}\left( {{{\mathcal K}_{ip}}} \right) + {\lambda _{\max }}\left( {{{\mathcal K}_{id}}} \right) \right]}{{\lambda _{\min }}\left( {{{\mathcal K}_{id}}} \right) - {\varrho _{2i}}},
    \end{array} \right.
\end{equation}
where $\kappa_3$ is defined right before Theorem \ref{t4}.
\hfill $\blacksquare$
\end{thm}

{Proof.}\,\,
First we proof
$\mathop {\lim }\nolimits_{t \to \infty } \left\| {\bar \varepsilon} \right\|_\infty \leq h \kappa_3$
and $\mathop {\lim }\nolimits_{t \to \infty } \left\| {\bar \upsilon} \right\|_\infty \leq h \kappa_3$.
Let $\lambda$ be an eigenvalue of $\Gamma$.
By Schur's Formula,
$\det (\lambda I_{2mn} - \Gamma) = \prod\nolimits_{i = 1}^n \left[ \psi_i (\lambda) \right]^m$,
where $\psi_i (\lambda) = {\lambda ^2} + \left[ {(\alpha h + \beta )(1 - s_i) - 2} \right]\lambda  + 1
- \beta (1 - s_i)$.
By Lemma \ref{l1}, Assumption A\ref{a1} implies $\left| {s_i} \right| < 1$, which means $1 - {s_i} \neq 0$.
Applying the bilinear transformation $\lambda = ({z + 1})/({z - 1})$ to $\psi_i (\lambda)$ gives that
$\psi^\prime_i (z) = {(z - 1)^2} \psi_i [({z + 1})/({z - 1})] / [\alpha h(1 - s_i)]$.
Then $\psi^\prime_i (z) = {z^2} + 2\beta z/(\alpha h) + 4/[\alpha h(1 - {s_i})] - 2\beta /(\alpha h) - 1$.
By the stability criterion, $\psi_i (\lambda)$ is {Schur stable} if and only if $\psi^\prime_i (z)$ is Hurwitz stable.
Then following \cite{Parks},
$\psi^\prime_i (z)$ is Hurwitz stable if and only if $2\beta/(\alpha h) > 0$ and
$\beta^2 \left[ 2({\theta _i} - \beta )/(\alpha h) - 1 \right] - \vartheta _i^2 > 0$.
It follows that if (\ref{5.2}) holds, $\psi_i (\lambda)$ is Schur stable, $i.e.$, $\Gamma \in \Omega^{2mn \times 2mn}$.
Note that the estimator in (\ref{2.0}) is equivalent to
$x (t_{k+1}) = \Gamma ^k x (t_1) + \sum\nolimits_{i = 1}^{k} {\Gamma^{k - i} \Delta(i)}$.
By the similar analysis in Appendix \ref{appendix1}, we can easily conclude that
$\mathop {\lim }\nolimits_{t \to \infty } \left\| {\bar \varepsilon} \right\|_\infty \leq h \kappa_3$,
and $\mathop {\lim }\nolimits_{t \to \infty } \left\| {\bar \upsilon} \right\|_\infty \leq h \kappa_3$.
\par
\setlength{\parskip}{0\baselineskip}
For the second presentation, note that
$\mathop {\lim }\nolimits_{t \to \infty } {\mu _{3i}}\left( t \right) \leq h \kappa_3 \sqrt m \left[ {\lambda _{\max }}\left( {{{\mathcal K}_{ip}}} \right) + {\lambda _{\max }}\left( {{{\mathcal K}_{id}}} \right) \right]$.
Following the proof of Theorem \ref{t2},
if ${\lambda _{\min }}\left( {{\mathcal{K}_{ip}}} \right) > \max \left( 2 \lambda_{iM} ; \varrho _{1i} \right)$
and ${\lambda _{\min }}\left( {{{\mathcal K}_{id}}} \right) > {\varrho _{2i}}$,
then (\ref{1.8}) implies
$\mathop {\lim }\nolimits_{t \to \infty } {\left\| {{e_i}\left( t \right)} \right\|_2} \leq \delta_5$,
$\mathop {\lim }\nolimits_{t \to \infty } {\left\| {{{\dot e}_i}\left( t \right)} \right\|_2} \leq \delta_6$.
This completes the proof.
\hfill $\blacksquare$
\par
\setlength{\parskip}{1\baselineskip}
Suppose that (\ref{5.2}) holds.
For any $\delta _5,\delta _6 > 0$, by Theorem \ref{t4}, if
there exists a constant $\epsilon \in (0,1)$ such that
\begin{equation}\label{6.2}
 \left\{ \begin{array}{lll}
  \lambda _{\min }\left( {{{\mathcal K}_{ip}}} \right) \geq \max \left( 2 \lambda_{iM} ;  \frac{\mu_{1i}}{\epsilon\delta _5} + {\varrho _{1i}} \right), \\
  \lambda _{\min }\left( {{{\mathcal K}_{id}}} \right) \geq \frac{\mu_{1i}}{\epsilon\delta _6} + {\varrho _{2i}}, \\
  h \leq \frac{\mu_{1i} (1 - \epsilon) }
  {\epsilon \kappa_3 \sqrt m \left[ {\lambda _{\max }}\left( {{{\mathcal K}_{ip}}} \right) + {\lambda _{\max }}\left( {{{\mathcal K}_{id}}} \right) \right]},
  \end{array} \right.
\end{equation}
then
$\mathop {\lim }\nolimits_{t \to \infty } {q_i} \in {\mathcal U}\left( {{\varepsilon_0};{\delta _5}} \right),~
\mathop {\lim }\nolimits_{t \to \infty } {\dot{q}_i} \in {\mathcal U}\left( {{\upsilon_0};{\delta _6}} \right).$

\begin{remark}
$\delta _5$ and $\delta _6$ in (\ref{5.3}) can be arbitrarily small
by choosing appropriate parameters $\mathcal{K}_{ip}, \mathcal{K}_{id}, h$
without the exact knowledge of ${\mu _{1i}}$, ${\mu _{2i}}$, ${\varrho _{1i}}$ and ${\varrho _{2i}}$.
The parameter setting follows the similar discussion in Remark \ref{remark1}.
\end{remark}

\begin{remark}
By Theorems \ref{t1} and \ref{t4},
the stability conditions of the first- and second-order DCEA (\ref{3.3})
and (\ref{2.2}) are (\ref{0.8}) and (\ref{5.2}), respectively.
Note that (\ref{5.2}) for the second-order DCEA restrict the upper bound of $h$ while (\ref{0.8}) dose not impose any restrictions on the sampling period $h$.
\end{remark}

\begin{remark}
The results in Theorems \ref{t1} and \ref{t2} for the first-order DCEA (\ref{3.3}) can be directly invoked to solve the sampled hetero-information case presented in \cite{Guan04} while the second-order DCEA (\ref{2.2}) in Theorem \ref{t4} cannot handle with the sampled hetero-information case directly.
\end{remark}

\begin{remark}
Comparing with the traditional estimator-based coordination algorithms considering continuous interaction \cite{fmg05,Mei01,Meng01},
the presented DCEA only transmit and update the interaction information
at the sampling time, which lead to the following benefits:
less requirement of target information (only use sampling data of the target),
lower cost for maintaining interaction (only require sampled interaction),
fewer consumption of calculation resources (only update the estimate value at sampling time).
\end{remark}

\begin{remark}
The traditional estimator-based coordination algorithms for NRSs based on continuous interaction
may be still effective for sampled interaction where the sampling period is sufficiently small.
However, the quantitative conditions and relationships between the sampling period,
feedback gains and the bound of the stability region cannot be obtained by the existing results
on traditional estimator-based algorithms.
\end{remark}

\begin{remark}
Comparing with the model-based adaptive algorithms of NRSs \cite{fmg01,fmg02,Wang01,Dong01},
the presented DCEA use model-free PD-like control,
which means the algorithms are easily realizable due to their low computation complexity, including
less requirement of the information of system models,
lower cost for real-time input computing.
\end{remark}

\section{Illustrative Examples}

Consider a NRS containing $6$ robots. Then ${\mathcal I} = \{ 1,\cdots,6\}$. The NRS interaction is described by a digraph $\Im$
with an adjacency matrix $\mathcal{W} = [0\;{\mathbf 0};\zeta\;{\hat{\mathcal W}}]$,
where $\zeta = {\rm col} (0,0,1,0,0,0)$ and
\begin{equation*}
  {\hat{\mathcal W}} =
  \left[ \begin{array}{*{20}{lll}}
         {0}&{1}&{1}&{0}&{0}&{0}\\
         {1}&{0}&{0}&{0}&{0}&{0}\\
         {0}&{0}&{0}&{0}&{0}&{0}\\
         {1}&{0}&{1}&{0}&{0}&{0}\\
         {0}&{1}&{1}&{1}&{0}&{0}\\
         {0}&{0}&{1}&{0}&{0}&{0}
        \end{array} \right]
  \in {\mathbb R}^{6 \times 6}.
\end{equation*}
For simplicity, the detailed setting of ${{\mathcal M}_i}\left( {q_i} \right)$, ${{\mathcal C}_i}\left( {{\dot q}_i},{q_i} \right)$ and ${{\mathcal G}_i}\left( {q_i} \right)$ in (\ref{0.1}) follows the dynamics of the robotic manipulator with two revolute joints given in \cite{Feng02}.
Let ${\tau _{i,d}}(t) = 2{\rm col}[\sin (t),\cos(2t)]$,
$\varepsilon_0(t) = {\rm col}[2t + \sin (t), -2t-\cos (t)]$ and
$\upsilon_0(t) = {\rm col}[2 + \cos (t), -2+\sin (t)]$, $\forall i \in {\mathcal I}$.
The elements of $\varepsilon_i(0)$, $\upsilon_i(0)$, $q_i(0)$ and $\dot q_i(0)$
are randomly selected from $[-25,25]$, ${\mathcal K}_{i,p} = 200 I_2$, ${\mathcal K}_{i,d} = 300 I_2$, $\forall i \in {\mathcal I}$.
\par
\begin{figure*}
  \centering
  \includegraphics[width=11cm]{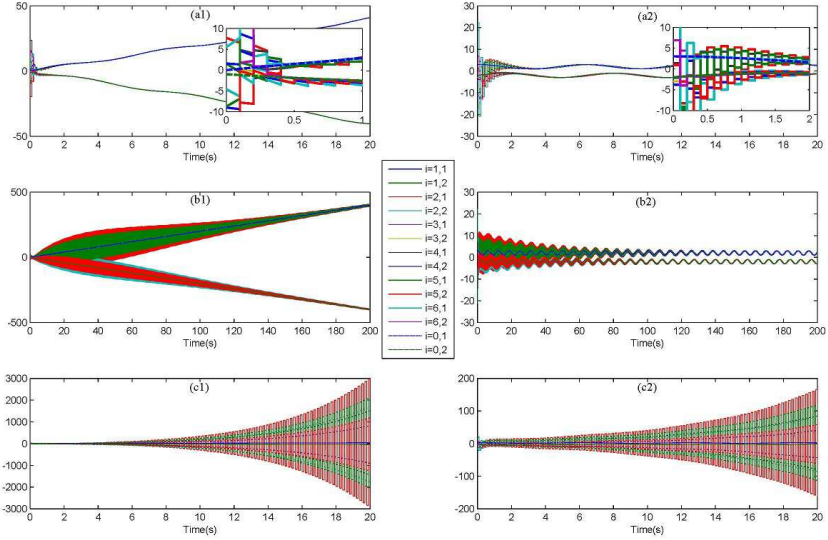}\\
  \caption{States $\varepsilon_i$ [Pictures (a1,b1,c1)] and $\upsilon_i$ [Pictures (a2,b2,c2)]
  of the first-order DCEA (\ref{3.3}) under the sampling period $h = 0.1$ and different $\alpha$ and $\beta$.
  (a1,a2) $\alpha = 0.9, \beta = 1.1$; (b1,b2) $\alpha,\beta = 1.17$; (c1,c2) $\alpha,\beta = 1.18$.}\label{Fig.e1}
\end{figure*}
\begin{figure*}
  \centering
  \includegraphics[width=11cm]{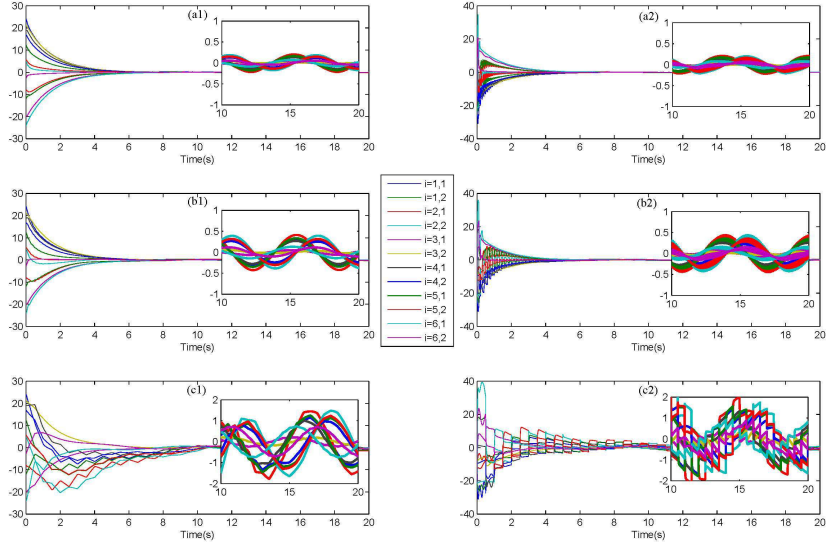}\\
  \caption{States $e_i$ [Pictures (a1,b1,c1)] and $\dot e_i$ [Pictures (a2,b2,c2)] of NRS (\ref{0.1}) using the first-order DCEA (\ref{3.3}) under $\alpha = 0.9$, $\beta = 1.1$ and different sampling periods $h$, $\forall i \in {\mathcal I}$.
  (a1,a2) $h = 0.05$; (b1,b2) $h = 0.1$; (c1,c2) $h = 0.5$.}\label{Fig.e2}
\end{figure*}
\begin{figure*}
  \centering
  \includegraphics[width=11cm]{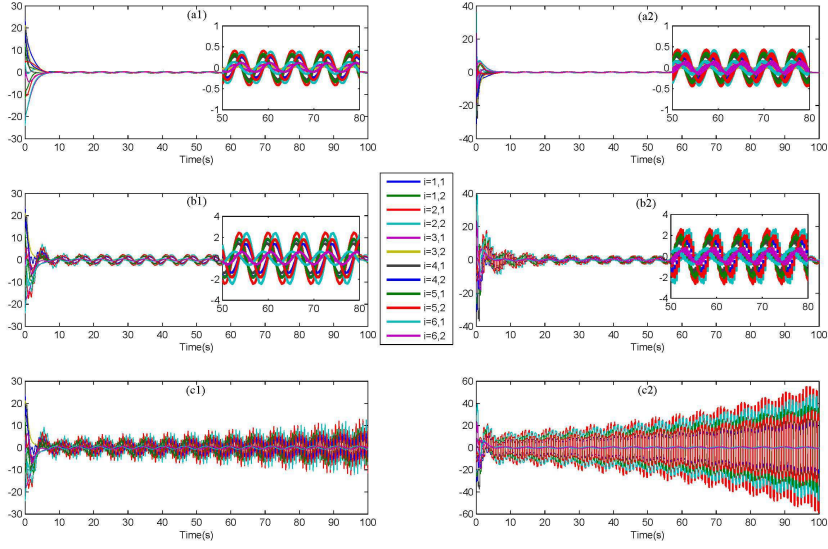}\\
  \caption{States $e_i$ [Pictures (a1,b1,c1)] and $\dot e_i$ [Pictures (a2,b2,c2)] of NRS (\ref{0.1}) using the second-order DCEA (\ref{2.2}) under $\alpha = 1.1$, $\beta = 0.9$ and different sampling periods $h$, $\forall i \in {\mathcal I}$.
  (a1,a2) $h = 0.1$; (b1,b2) $h = 0.45$; (c1,c2) $h = 0.5$.}\label{Fig.e4}
\end{figure*}

\begin{example}
Let $h = 0.1$. For the first-order DCEA (\ref{3.3}),
the practical stability of $\varepsilon_i$ and $\upsilon_i$ can be achieved if $0 < \alpha,\beta < 1.1716$,
which can be easily computed from (\ref{0.8}) in Theorem \ref{t1}.
Fig.\ref{Fig.e1} shows that the practical stability of $\varepsilon_i$ and $\upsilon_i$ can be achieved
when $\alpha = 0.9, \beta = 1.1$ and $\alpha, \beta = 1.17$, but it cannot be achieved if $\alpha, \beta = 1.18$.
\end{example}
\begin{example}
Let $\alpha = 0.9$ and $\beta = 1.1$.
The detail views in Fig.\ref{Fig.e2} are the amplifications of steady-state errors for NRS (\ref{0.1}) under DCEA (\ref{3.3}).
It is shown in Fig.\ref{Fig.e2} that the smaller the sampling period $h$, the smaller the stability region.
\end{example}
\begin{example}
Theorem \ref{t4} implies that for NRS (\ref{0.1}) under DCEA (\ref{2.2}), the practical stability of the NRS can be achieved if $\beta < 1.1716$.
Let $\alpha = 1.1$ and $\beta = 0.9$.
By (\ref{5.2}), target tracking can be achieved if $0 < h < 0.4938$.
Fig.\ref{Fig.e4} shows that the target tracking can be achieved
when $h = 0.45$, but it cannot be obtained when $h = 0.5$.
It follows that the sufficient conditions (\ref{0.8}) and (\ref{5.2}) are almost equivalent to necessary and sufficient conditions.
Besides, it can be seen that in some sense, the smaller the sampling period $h$, the smaller the stability region.
\end{example}

\begin{figure*}
  \centering
  \includegraphics[width=11cm]{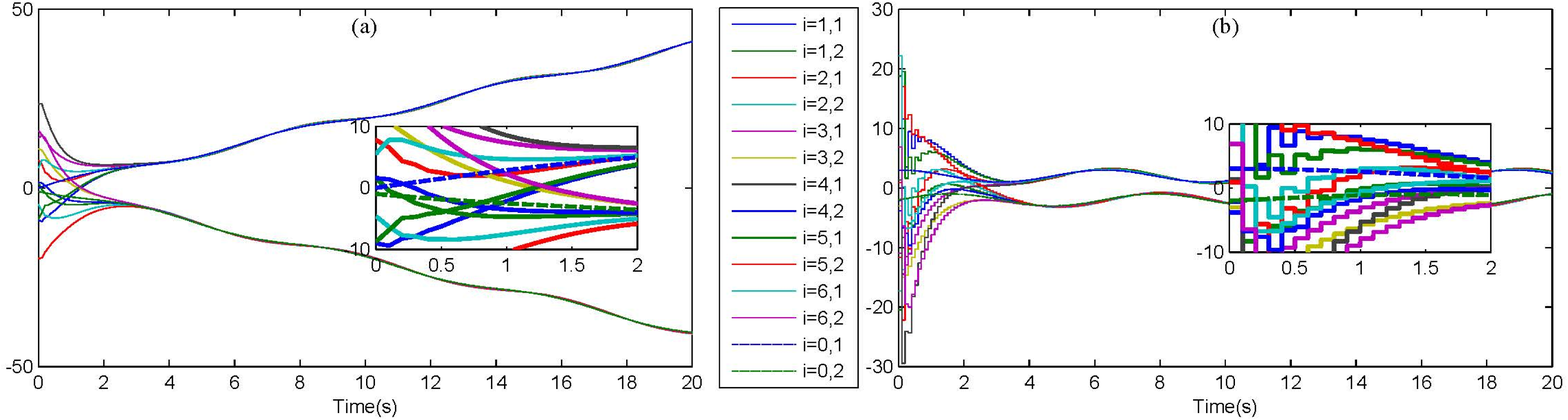}\\
  \caption{States $\varepsilon_i$ [Picture (a)] and $\upsilon_i$ [Picture (b)] of the second-order DCEA (\ref{2.2}) under $h = 0.1$, $\alpha = 1.1$, and $\beta = 0.9$, $\forall i \in {\mathcal I}$.}\label{Fig.e3}
\end{figure*}

\begin{remark}
By the detail views in Fig.\ref{Fig.e1} and Fig.\ref{Fig.e3}, $\varepsilon_i$
is discontinuous at each sampling time in the first-order DCEA while it is continuous in the second-order DCEA,
which means that the trajectory of $\varepsilon_i$ and $q_i$ for NRS under the second-order DCEA is smoother
than that under the first-order DCEA.
\end{remark}

\section{Conclusion}

In this paper, the target tracking problem of NRSs under directed sampled interaction
has been considered for a time-varying target.
The concepts and properties of the small-value norms for Schur stable matrices
have been introduced for the analysis of the practical stability of the presented first- and second-order DCEA.
Several sufficient criteria on interaction topology, the sampling period,
and the other control parameters for target tracking have been obtained.
Besides, the quantitative relationship between the stability region of the tracking errors and the control parameters,
including the sampling period, the control gain and the interaction topology, has been presented.
Finally, a few examples have been delivered to verify the theoretical results.


\appendix
\section{Proof of Theorem \ref{t1}}\label{appendix1}    

For the first presentation, let $\lambda$ be an eigenvalue of $\Lambda$.
By Schur's Formula,
$\det (\lambda I_{2mn} - \Lambda) = \prod\nolimits_{i = 1}^{n} \left[ \varphi_i (\lambda) \right]^m$,
where $\varphi_i (\lambda) = (\lambda + \alpha -1 - \alpha {s_i})(\lambda + \beta -1 - \beta {s_i})$.
Then the eigenvalues of $\Lambda$ satisfy $\varphi_i (\lambda) = 0$.
Invoking the bilinear transformation $\lambda = {(z+1)}/{(z-1)}$ for $\varphi_i (\lambda)$ gives
$\varphi_i^\prime (z) = ({z-1})^2 \varphi_i \left[ {(z+1)}/{(z-1)} \right]$.
Then $\varphi_i^\prime (z) = [{\alpha (1 - {s_i})z - \alpha (1 - {s_i}) + 2}] [{\beta (1 - {s_i})z - \beta (1 - {s_i}) + 2}]$.
By invoking Lemma \ref{l1}, Assumption A\ref{a1} implies $1 - {s_i} \neq 0$.
$\varphi_i^\prime (z) = 0$ implies $z_1 = 1 - 2/(\alpha [1 - {s_i}])$ and $z_2 = 1 - 2/(\beta [1 - {s_i}])$.
It follows that ${\rm Re}(z_1) < 0$ and ${\rm Re}(z_2) < 0$ if (\ref{0.8}) holds, $i.e.$,
$\varphi_i^\prime (z)$ is Hurwitz stable if (\ref{0.8}) holds.
By stability criterion, $\varphi_i (\lambda)$ is {Schur stable} if and only if polynomial $\varphi_i^\prime (z)$ is {Hurwitz stable}.
Therefore, (\ref{0.8}) implies that $\Lambda$ has all eigenvalues within the unit disc,
$i.e.$, $\Lambda \in \Omega^{2mn \times 2mn}$.
\par
\setlength{\parskip}{0\baselineskip}
For the second presentation, Assumption A\ref{a2}
means that for any $k \in \mathbb{Z}^\dag$,
\begin{equation}\label{0.7}
\left\{ \begin{array}{lll}
  \left\| \Delta_1 \left( k \right) \right\|_\infty \leq 2h \gamma_1, ~
  \left\| \Delta_2 \left( k \right) \right\|_\infty \leq h \gamma_2, \\
  \left\| \Delta \left( k \right) \right\|_\Lambda \leq h c_1 \max(2 \gamma_1 ; \gamma_2)
  \end{array} \right.
\end{equation}
where we have invoked ${\varepsilon _0}(t_{k+1}) - {\varepsilon _0}({t_k}) = \int_{t_k}^{t_{k+1}} {v(w)dw}$ and
${{\upsilon}_0}(t_{k+1}) - {{\upsilon}_0}(t_{k}) = \int_{t_k}^{t_{k+1}} {a(w)dw}$ to obtain (\ref{0.7}).
Following \cite{Horn}, $\Lambda \in \Omega^{2mn \times 2mn}$ means
$\mathop {\lim }\nolimits_{k \to \infty } {\Lambda ^k} = 0$.
System (\ref{2.3}) gives that
$x (t_{k+1}) = \Lambda ^k x (t_1) + \sum\nolimits_{i = 1}^{k} {\Lambda^{k - i} \Delta(i)},$
where $k \in \mathbb{Z}^\dag$, $\Lambda^0 = I_{2mn}$.
It thus follows from $\mathop {\lim }\nolimits_{k \to \infty } {\Lambda ^k} = 0$ that
\begin{equation*}
\begin{array}{lll}
    \mathop {\lim }\limits_{t \to \infty } \left\| x (t) \right\|_\Lambda
    &=& \mathop {\lim }\limits_{k \to \infty } \left\| \sum\limits_{i = 1}^{k} {\Lambda^{k - i} \Delta(i)} \right\|_\Lambda \\
    &\leq& h c_1 \max(2 \gamma_1 ; \gamma_2) \mathop {\lim }\limits_{k \to \infty }  \sum\limits_{i = 0}^{k-1} \left\| \Lambda \right\|_\Lambda^i   \\
    &\leq& h c_1 \max(2 \gamma_1 ; \gamma_2) (1 - \left\| \Lambda \right\|_\Lambda)^{-1},
    \end{array}
\end{equation*}
where Lemma \ref{l3} has been used to obtain the above result.
It follows that $\mathop {\lim }\nolimits_{t \to \infty } \left\| x (t) \right\|_\infty \leq h \kappa_1$,
which means $\mathop {\lim }\nolimits_{t \to \infty } \left\| {\bar \varepsilon}_i \right\|_\infty \leq h \kappa_1$
and $\mathop {\lim }\nolimits_{t \to \infty } \left\| {\bar \upsilon}_i \right\|_\infty \leq h \kappa_1$, $\forall i \in {\mathcal I}$.
Besides, the estimator in (\ref{2.3}) also implies that
$\bar{\upsilon} (t_{k+1}) = {\mathcal P}^k \bar{\upsilon} (t_1) + \sum\nolimits_{i = 1}^{k} {{\mathcal P}^{k - i} \Delta_2 \left( i \right)}$, $\forall k \in \mathbb{Z}^\dag$,
where ${\mathcal P}$ is defined right behind (\ref{0.5}).
Thus, by the similar analysis, ${\mathcal P} \in \Omega^{mn \times mn}$ and
$\mathop {\lim }\nolimits_{t \to \infty } \left\| \bar \upsilon _i \right\|_\infty \leq h \kappa_2$,
$\forall i \in {\mathcal I}$.
This completes the proof.
\hfill $\blacksquare$

\setlength{\parskip}{1\baselineskip}

\end{document}